\documentclass[aps, prl, reprint, showpacs]{revtex4-1}

\usepackage{amsfonts}
\usepackage{amsmath}
\usepackage{amssymb, bm, mathrsfs}
\usepackage{graphicx, epstopdf, color}
\usepackage{soul}
\usepackage[colorlinks=true,citecolor=blue]{hyperref}

\begin{document}

\title{Universal Scaling Laws in Schottky Heterostructures Based \\ on Two-Dimensional Materials }

\author{Yee Sin Ang}
\email{yeesin\_ang@sutd.edu.sg}

\author{Hui Ying Yang}

\author{L. K. Ang}
\email{ricky\_ang@sutd.edu.sg}

\affiliation{SUTD-MIT International Design Center \& Science and Math Cluster, Singapore University of Technology and Design (SUTD), 8 Somapah Road, Singapore 487372}

\begin{abstract}

	We identify a new universality in the carrier transport of two-dimensional(2D)-material-based Schottky heterostructures. We show that the reversed saturation current ($\mathcal{J}$) scales \emph{universally} with temperature ($T$) as $ \log(\mathcal{J}/T^{\beta}) \propto -1/T$, with $\beta = 3/2$ for lateral Schottky heterostructures and $\beta = 1$ for vertical Schottky heterostructures, over a wide range of 2D systems including nonrelativistic electron gas, Rashba spintronic system, single and few-layer graphene, transition metal dichalcogenides and thin-films of topological solids. 
	Such universalities originate from the strong coupling between the thermionic process and the in-plane carrier dynamics.
	Our model resolves some of the conflicting results from prior works and is in agreement with recent experiments. 
	The universal scaling laws signal the breakdown of $\beta=2$ scaling in the classic diode equation widely-used over the past 60 years. Our findings shall provide a simple analytical scaling for the extraction of the Schottky barrier height in 2D-material-based heterostructure, thus paving way for both fundamental understanding of nanoscale interface physics and applied device engineering. 
	
\end{abstract}

\maketitle

Contacting two-dimensional (2D) material with a bulk material or another 2D material to form a heterostructure \cite{allain, xu} is an inevitable process for nanoelectronics \cite{fiori} and optoelectronics \cite{koppens}.
The contact often leads to the formation of an interface barrier (or \emph{Schottky barrier}) with a Schottky barrier height (SBH) denoted by $\Phi_{B}$.
For SBH significantly lower than the thermal energy ($\Phi_{B} \ll k_BT$), the heterostructure becomes a non-rectifying \emph{Ohmic contact}. 
Otherwise, a rectifying junction, termed as \emph{Schottky contact}, is formed.
2D-material-based Schottky heterostructures have been actively studied in recent years \cite{x_wu, cronin, tongay, x_wu, sinha, liang, ang, ang_MRS, bartolomeo2, tomer, NR, 2DM, h_yu, wchen, marian, grassi, somvanshi, yang_bar, trushin} due to their broad applications such as transistor \cite{yang_bar}, photodetection \cite{sanchez}, energy harvesting \cite{li_solar}, sensing \cite{kim_sensor} and data storage \cite{d_li}.

For an ideal Schottky heterostructure under a bias voltage $V$, the electrical current density $J$ is governed by the Shockley diode equation, $J = \mathcal{J}[\exp{(eV/k_BT)} - 1]$ \cite{shockley}.
Here, the \emph{reversed saturation current density} (RSC), $\mathcal{J}$, originates from the thermionic electron emission over the Schottky barrier \cite{RD} at reverse-bias [see Fig. 1(a)]. The $\mathcal{J}$ can be expressed as a \emph{generalized Richardson formula}:
\begin{equation}\label{sr}
\log\left(\frac{\mathcal{J}}{T^{\beta}}\right) = \mathcal{A} - \frac{ \mathcal{B} }{T} ,
\end{equation}
where $\mathcal{A}$ and $\mathcal{B}$ are material/interface-dependent constants. 
Equation (\ref{sr}) is a universal hallmark of the thermionic transport and the scaling exponent, $\beta$, takes the Richardson-Dushman form of $\beta = 2$ for Schottky contact formed by three-dimensional (3D) bulk metals with parabolic energy dispersion \cite{RD}.
Importantly, Eq. (\ref{sr}) together with the known $\beta$ provide a simple tool for the extraction of SBH. It contains great wealth of interface physics and is critical to the operation and performance of all Schottky-contact-based functional devices \cite{tung}.

For 2D-material-based Schottky heterostructures, the accurate extraction of SBH is particularly important as the SBH can exhibit its complex dependences on factors such as lattice mismatch \cite{farmanbar}, strain \cite{strain}, metal work function \cite{farmanbar2}, layer thickness \cite{kwon}, electric-field effect \cite{yang_bar, bar} and so on. 
Thus, analytical transport models, akin to Eq. (\ref{sr}), for 2D-material-based Schottky heterostructure are highly valuable for both fundamental interface physics and device engineering.

\begin{figure}[t]
	\includegraphics[scale = 0.423]{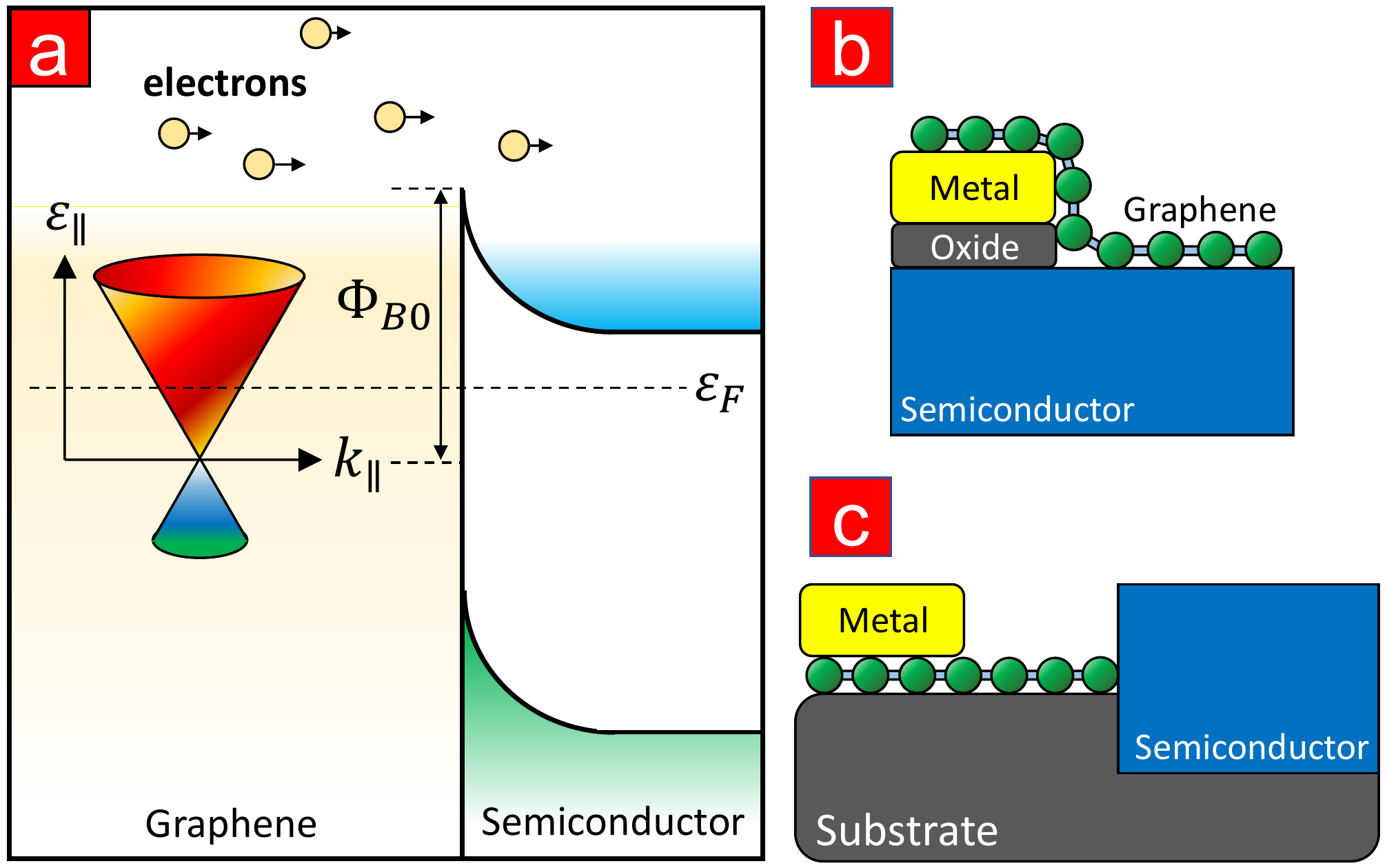}
	\caption{(a) Band diagram of a graphene-based Schottky heterostructure showing the thermionic transport over a Schottky barrier ($\Phi_{B0}$). Schematic drawing of a graphene-based (b) vertical; and (c) lateral 2D/3D Schottky heterostructure.}
\end{figure}

Despite enormous efforts devoted to the study of 2D-material-based Schottky heterostructures, several inconsistencies and confusions regarding the carrier transport physics still persist in the literatures.
Taking graphene-based Schottky heterostructure as an example, various practices of fitting the RSC data with different $\beta$ confusingly co-exist in the literatures \cite{x_wu, cronin, tongay, sinha, liang, ang, ang_MRS, bartolomeo2, tomer, NR, 2DM}. 
This inconsistency is further complicated by the presence of two distinct contact configurations, namely the \emph{vertical Schottky heterostructure} (VSH) and the \emph{lateral Schottky heterostructure} (LSH), which may exhibit completely different $\beta$-scaling [see Figs. 1(b) and (c)].
Although recent works have shed light on the growth \cite{x_chen}, structural \cite{mcguigan_thick}, thermal \cite{x_liu}, electrostatic \cite{zheng}, electronic \cite{w_chen} and electrical \cite{zhao, ling, guimaraes, behranginia} properties of 2D-material-based LSH, a consistent model remains lacking.

In this paper, we develop generalized analytical RSC models for LSH and VSH over a wide range of 2D electronic systems, including nonrelativistic electron gas, Rashba spintronic system, single and few-layer graphene, and thin-films of topological solids.
The key findings are of three-fold.
Firstly, we identify a \emph{universal current-temperature scaling exponent} of $\beta = 3/2$ in LSH, which is independent on the types of 2D electronic systems.
Secondly, for VSH with non-conserving lateral electron momentum induced by carrier scattering effects \cite{meshkov, russell}, we report another universal scaling exponent of $\beta = 1$.
These scaling universalities are absent in the bulk-material-based Schottky heterostructure and are in unison with multiple experimental results.
Thirdly, for graphene-based VSH, we unify the two prior contrasting models of $\beta = 1$ due to Sinha and Lee \cite{sinha}, and of $\beta = 3$ due to Liang and Ang \cite{liang} under the physical framework of lateral electron momentum conservation \cite{meshkov,russell, vashaee}. 
An important consequence of our results is that the classic diode scaling of $\beta = 2$, a cornerstone theory for understanding bulk diode transport physics over the past 60 years \cite{p_zhang}, is no longer valid for 2D materials. A timely paradigm shift from the classic $\beta = 2$ scaling to the new scaling laws developed here is required to better capture the interface physics of 2D material heterostructures as required in many applications.

We consider a 2D nanosheet lies in the $x$-$y$ plane and in contact with a bulk/2D-semiconductor via its edge. The RSC flowing across the heterostructure is
\begin{equation} \label{edge_J_gen}
\mathcal{J}( k_F, T ) = \frac{g_{s,v} e }{(2\pi)^2} \sum_{k_\perp^{(i)}}\int \text{d}^2\mathbf{k}_\parallel v_x (k_\parallel) f(\mathbf{k}_\parallel, k_F ) \mathcal{T}(k_x) ,
\end{equation}
where $\mathbf{k}_\parallel = (k_x, k_y)$ is the electron wave vector lying in the plane of 2D material, $\mathcal{T}(k_x)$ is the transmission probability, $v_x (k_x) = \hbar^{-1} \partial \varepsilon_\parallel / \partial k_x = (\hbar^{-1} \partial \varepsilon_\parallel / \partial |\mathbf{k}_\parallel| ) \cos \phi $ is the $x$-directional group velocity, $f(\mathbf{k}_\parallel, k_F)$ is the carrier distribution function, $k_F$ is the Fermi wave vector, $\phi = \tan^{-1} (k_y/k_x)$ and $g_{s,v}$ is the spin-valley degeneracy. For over-barrier emission, $\mathcal{T}(k_x) = \Theta\left[k_x(\varepsilon_\parallel) - |\mathbf{k}_\parallel(\Phi_{B0})|\right]$, where $\Theta(x)$ is a Heaviside step-function and $\Phi_{B0}$ is the SBH measured from zero-energy [see Fig. 1(a)].
As the material thickness increases, $k_\perp^{(i)}$'s becomes closely-spaced, and the $\sum_{k_\perp^{(i)}} \to (L_\perp / 2\pi) \int dk_\perp $ transforms Eq. (\ref{edge_J_gen}) into the 3D counterpart.

\begin{table*}[!t]
	\caption{Universal scaling exponent ($\beta = 3/2$) for lateral Schottky heterostructures made of various 2D systems: nonparabolic 2D electron gas ($\gamma$-2DEG), Rashba spintronic system (R-2DEG), gapless and gapped Dirac cones (Dirac), metallic transition metal dichalcogenides (TMD) with 2H structural phase where the valence Fermi pockets are composed of parabolic dispersion at $\Gamma$ point and gapped Dirac dispersion at $K$ and $K'$ points, $ABA$-stacked few-layer-graphene ($ABA$-FLG) and $ABC$-stacked few-layer-graphene ($ABC$-FLG).  Here $\gamma$ is the band nonparabolicity parameter, $s=\pm1$ denotes the two Rashba spin-split subbands, $\Lambda_\pm \equiv \sqrt{ 1 \pm (1 + \Phi_{B0}/\varepsilon_R)^{-1/2} }$, $\varepsilon_R \equiv m^*\alpha_R^2/2\hbar^2$ is the Rashba parameter, $\alpha_R$ is the Rashba spin-orbit coupling strength, $v_F$ is the Fermi velocity of Dirac dispersion, $\Delta$ is the band gap, $\alpha_n^{(N)} = t_\perp \cos(n\pi/(N+1) )$, $t_\perp =0.39$ eV, $n = 1,2, \cdots, N$ denotes the $n$-th subband and $N$ is the layer number. Note that $N\geq 3$ and $N\geq1$ for $ABA$-and $ABC$-FLG, respectively. }
	\includegraphics[scale = 1]{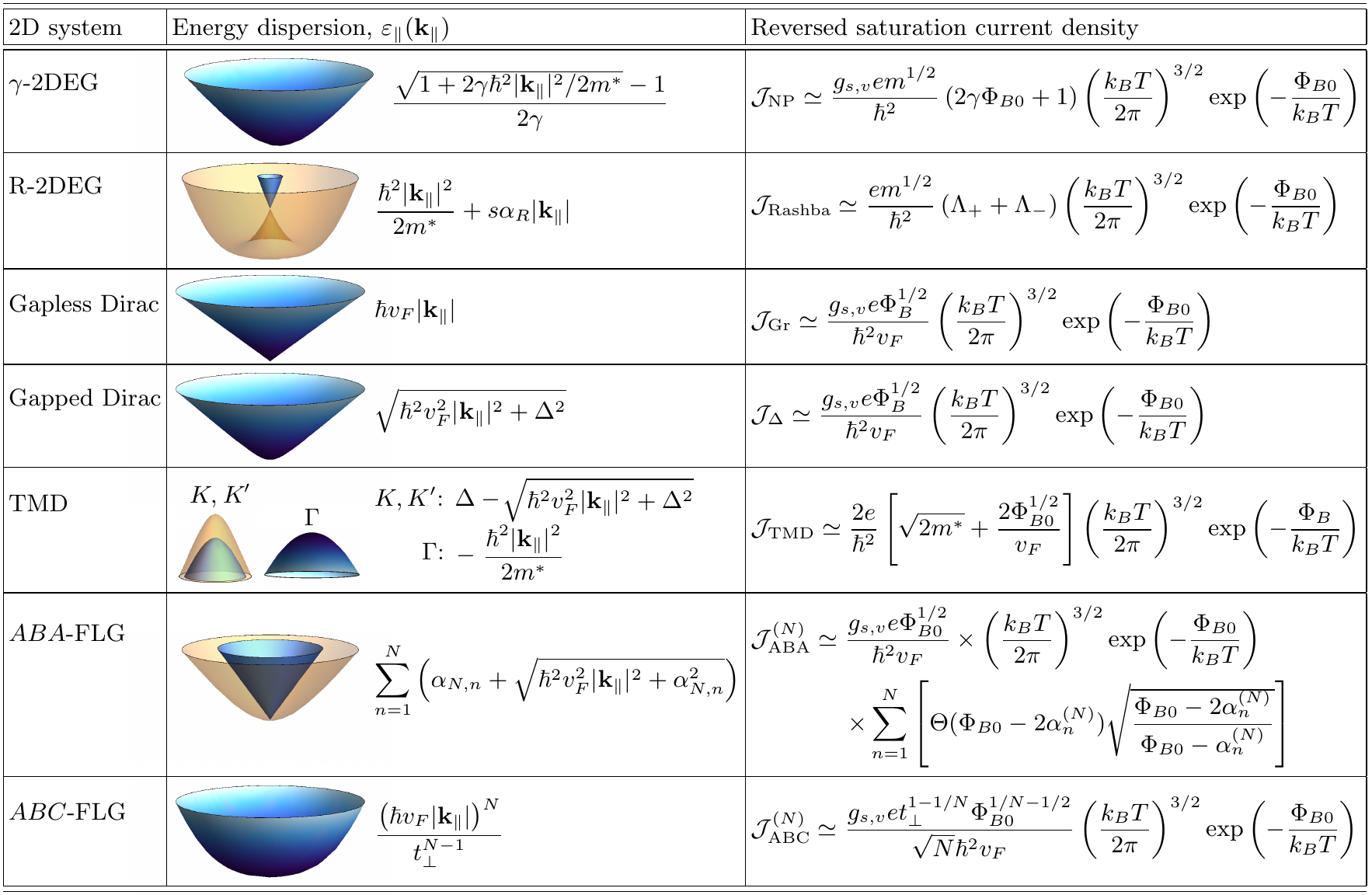}
\end{table*}

Consider a 2D material with a general polynomial form of isotropic energy dispersion,
\begin{equation} \label{poly}
\varepsilon_\parallel(\mathbf{k}_\parallel) = \sum_n c_n |\mathbf{k}_\parallel|^n ,
\end{equation} 
where $c_n$ is a coefficient, and $n \in \mathbb{Z}^{\geq}$. 
The explicit solution of $|\mathbf{k}_\parallel|$, to be solved from Eq. (\ref{poly}), is not required to derive the current-temperature scaling relation.
We only need to express $|\mathbf{k}_\parallel|$ as
\begin{equation}\label{k_sol}
\left|\mathbf{k}_\parallel (\varepsilon_\parallel) \right|= \sum_l \alpha_l \varepsilon_\parallel^{l},
\end{equation}
where $l  \in \mathbb{Z}^{\geq}$ and $\alpha_l$ are terms dependent on the explicit form of Eq. (\ref{poly}).
The $\mathcal{T}(k_x)$ can be decomposed as
\begin{eqnarray}\label{T}
\mathcal{T}(k_x) &=&  \Theta \left( X - \frac{ \left |\mathbf{k}_\parallel (\Phi_{B0}) \right| }{ \left| \mathbf{k}_\parallel  (\varepsilon_\parallel)\right|} \right)\Theta \left( \left| \mathbf{k}_\parallel (\varepsilon_\parallel) \right| - \left| \mathbf{k}_\parallel(\Phi_{B0}) \right| \right) , \nonumber \\
\end{eqnarray}
where $X \equiv \cos \phi$. The second step-function ensures that $ |\mathbf{k}_\parallel(\Phi_{B0})| / |\mathbf{k}_\parallel(\varepsilon_\parallel)| \leq 1$ since $X$ cannot exceed unity for any real-valued $\phi$.
By using $v_x(\varepsilon_\parallel) |\mathbf{k}_\parallel| d |\mathbf{k}_\parallel| = \hbar^{-1} X |\mathbf{k}_\parallel| d\varepsilon_\parallel $ and $X d\phi =dX / \sqrt{1 - X^2} $, Eqs. (\ref{T}) and (\ref{edge_J_gen}) are combined to give 
\begin{eqnarray} \label{J_int}
\mathcal{J}
&=& \frac{2g_{s,v} e }{(2\pi)^2 \hbar} \int_{0}^{\infty} d \varepsilon_\parallel \mathsf{K}_2^{1/2} \Theta \left( \mathsf{K}_1\right) f(\varepsilon_\parallel,\varepsilon_F),
\end{eqnarray}
where $\mathsf{K}_n \equiv \left|\mathbf{k}_\parallel(\varepsilon_\parallel) \right|^n- \left|\mathbf{k}_\parallel(\Phi_{B0}) \right|^n $. The $\mathsf{K}_2$ is solved as
\begin{eqnarray}\label{expand}
\mathsf{K}_2
&\simeq& \left[  \sum_l \alpha_l \Phi_{B0}^l  \left( 1 + l\frac{\mu}{u_0} \right) \right]^2 -  \left( \sum_l \alpha_l \Phi_{B0}^l \right)^2 \nonumber \\
&\simeq& 2 k_BT \mu  / \Phi_{B0} \times h_{ll'} ,
\end{eqnarray}
where  $h_{ll'}\equiv \sum_{ll'}  l \alpha_l \alpha_{l'} \Phi_{B0}^{l + l'} $, $\mu \equiv (\varepsilon_\parallel - \Phi_{B0}) / k_BT$ and $u_0 \equiv \Phi_{B0}/k_BT$. 
In the derivation of Eq. (\ref{expand}), we have used Eq. (\ref{k_sol}) and performed a Taylor expansion of $\mu \ll \mu_0$ (valid for thermionic emission in the non-degenerate regime).
By simplifying the step-function in Eq. (\ref{J_int}) as
\begin{eqnarray}\label{step}
\Theta \left( \mathsf{K}_1 \right) &\simeq& \Theta\left( \sum_l \alpha_l u_0^l \left[ \left( \frac{\mu}{u_0} + 1\right)^l - 1 \right] \right) \nonumber \\
&\simeq& \Theta\left( \sum_l l\alpha_l u_0^{l-1} \mu \right) = \Theta(\mu) ,
\end{eqnarray}
Eqs. (\ref{J_int}-\ref{step}) are jointly solved to obtain the central result of this work ($\Phi_{B} \equiv \Phi_{B0}-\varepsilon_F$) \cite{SM}:
\begin{equation}\label{J_solved}
\mathcal{J} \simeq  \frac{g_{s,v} e }{ \hbar  } \sqrt{\frac{h_{ll'}}{\Phi_{B0}}}\left( \frac{k_BT}{2\pi} \right)^{3/2} \exp\left( - \frac{\Phi_B}{k_BT} \right),
\end{equation}
which reveals a \emph{universal scaling law} of $\beta = 3/2$, i.e.
\begin{equation}\label{USR}
\log\left(\frac{\mathcal{J}}{T^{3/2}} \right) \propto -\frac{1}{T}.
\end{equation}
The physical origin of the $\beta = 3/2$ universal scaling relation can be understood as followed.
In bulk-material-based Schottky heterostructure, the strong coupling between the out-of-plane electron dynamics and the thermionic tunneling process pins the RSC into the generalized Richardson form of Eq. (\ref{sr}). 
Nonetheless, $\varepsilon_\parallel$ remains decoupled from the cross-plane transport process, which leaves the scaling exponent, $\beta$, variable depending on $\varepsilon_\parallel$. In contrary, the lateral contact geometry of LSH interlocks the in-plane $\varepsilon_\parallel$ with the in-plane thermionic tunneling process. Such interlocking pins the scaling exponent to the universal and material-independent value of $\beta = 3/2$ regardless the form of $\varepsilon_\parallel$. 

\begin{table*}[!t]
	\caption{Reversed saturation current density of vertical Schottky heterostructures with and without $\mathbf{k}_\parallel$-conservation. Here, $\xi_T \equiv \exp(-\Phi_B/k_BT)$, $\text{erf}(x) \equiv \pi^{-1/2} \int_{-x}^x \exp(-t^2) dt$ is the error function, $\Gamma(s,x) \equiv \int_x^{\infty} t^{s-1} \exp(-t) dt$ is the upper incomplete Gamma function, and $\Gamma(s) \equiv \int_0^{\infty} t^{s-1} \exp(-t) dt$ is the complete Gamma function. All $\mathcal{J}_\star^{(j=1)}$ converge to $\beta = 1$ for $\Phi_B \gg k_BT$. }
	\begin{tabular}{ l  l  l }
		\hline\hline
		2D system  & Conserving lateral momentum ($j=0$)  & Non-conserving lateral momentum ($j=1$)   \\ \hline  
		$\gamma$-2DEG & $\displaystyle\mathcal{J}_{\star\text{NP}}^{(j=0)} =  \frac{g_{s,v} e m^{*}k_B^2}{4\pi^2\hbar^3} \left[ T^2 + 2\gamma k_BT^3 \right] \xi_T$ & $\displaystyle\mathcal{J}_{\star \text{NP} }^{(j=1)} = \frac{\lambda g_{s,v} e m^* v_\perp}{2\pi \hbar^2 L_\perp} \left[ 1 +  2\gamma k_BT \left( 1 + \frac{\Phi_{B0}}{k_BT}\right) \right] k_BT\xi_T$  \\
		&& \\
		R-2DEG & $\displaystyle \mathcal{J}_{\star\text{Rashba}}^{(j=0)} = \frac{ em^*}{2\pi^2 \hbar^3} \left(k_BT\right)^2 $  & $ \displaystyle\mathcal{J}_{\star \text{Rashba}}^{(j=1)} =  \frac{\lambda e m^* v_\perp }{\pi \hbar^2 L_\perp} k_B T \xi_T $ \\ 
		& \hfil $ \displaystyle \times \left[ 1 + \sqrt{\frac{\pi \varepsilon_R}{k_BT}} \text{erf}\left( \sqrt{\frac{\varepsilon_R}{k_BT}} \right) \right] \xi_T $ & \\
		&& \\
		Gapless Dirac   & 	$ \displaystyle\mathcal{J}^{(j=0)}_{\star\text{Gr}} = \frac{ g_{s,v}e   }{4\pi^2 \hbar^3 v_F^2} \left(k_BT\right)^3 \xi_T $   &  $ \displaystyle	\mathcal{J}^{(j=1)}_{\star\text{Gr}} =  \frac{\lambda g_{s,v}e v_\perp}{2\pi \hbar^2 v_F^2 L_\perp} \left(k_BT\right)^2 \left( 1 + \frac{\Phi_{B0}}{k_BT} \right)  \xi_T$ \\
		&& \\
		Gapped Dirac   & $\displaystyle \mathcal{J}_{\star\Delta}^{(j=0)} = \frac{ g_{s,v}e   }{4\pi^2 \hbar^3 v_F^2} \left(k_BT\right)^3 \xi_T  $ & $ \displaystyle	\mathcal{J}^{(j=1)}_{\star \Delta} =  \frac{\lambda g_{s,v}e v_\perp}{2\pi \hbar^2 v_F^2 L_\perp} \left(k_BT\right)^2 \left( 1 + \frac{\Phi_{B0}}{k_BT} \right)  \xi_T$ \\
		TMD & $\displaystyle \mathcal{J}_{\text{TMD}\star}^{(j=0)} = \frac{em^*k_B^2}{2\pi^2 \hbar^3}\left( T^2 + \frac{k_B}{m^* v_F^2} T^3 \right) \xi_T$ & $\displaystyle \mathcal{J}_{\text{TMD}\star}^{(j=1)} = \frac{ \lambda e}{\pi \hbar^2}  \left[1  + \frac{2 k_BT}{m^*v_F^2}\left( 1 + \frac{\Phi_{B0}}{k_BT} \right) \right] k_BT \xi_T$ \\
		&& \\
		$ABA$-FLG & $\displaystyle\mathcal{J}^{(N, j=0)}_{\star\text{ABA}} = N\frac{g_{s,v}e}{4\pi^2\hbar^3 v_F^2} \left(k_BT\right)^3 \xi_T $  & $\displaystyle\mathcal{J}^{(N, j=1)}_{\star\text{ABA}} = N \frac{\lambda g_{s,v} e v_\perp}{2\pi \hbar^2 v_F^2 L_\perp} \left( k_BT \right)^2 \left( 1 + \frac{\Phi_{B0}}{k_BT} \right) \xi_T $ \\
		&& \\
		$ABC$-FLG & $\displaystyle \mathcal{J}_{\star\text{ABC}}^{(N,j=0)} = \frac{\Gamma(2/N)}{N}  \frac{g_{s,v}e t_\perp^{2-2/N} }{4\pi^2\hbar^3 v_F^2} \left(k_BT\right)^{2/N+1} \xi_T $  &  $\displaystyle \mathcal{J}^{(N, j=1)}_{\star \text{ABC}} =  \frac{1}{N} \frac{\lambda g_{s.v}e t_\perp^{2-2/N} \tilde{v}_\perp }{2\pi \hbar^2 v_F^2 L_\perp} \left( k_BT \right)^{2/N} \Gamma\left( \frac{2}{N}, \frac{\Phi_{B0}}{k_BT} \right) e^{\frac{\Phi_{B0}}{k_BT} } \xi_T $ \\ 
		&& \hfil $\displaystyle \simeq \frac{1}{N} \frac{\lambda g_{s.v}e \tilde{v}_\perp \Phi_{B0}^{2/N-1} }{2\pi \hbar^2 v_F^2 L_\perp t_\perp^{2/N - 2} } k_BT\xi_T $ \\ \hline \hline
	\end{tabular}
\end{table*}

We explicitly solve the RSC for a large variety of 2D electronic systems, including: 
(i) nonrelativistic 2D electron gas \cite{anwar} with band nonparabolicity correction ($\gamma$-2DEG) \cite{kane}; (ii) 2D Rashba spintronic system (R-2DEG) \cite{rashba}; (iii) gapless Dirac cone in honeycomb lattices such as graphene, silicene, germanene and stanene \cite{xene}; (iv) gapped Dirac cone in honeycomb lattices with broken inversion symmetry \cite{gap} and in the thin films of topological insulator and Dirac/Weyl semimetal \cite{TI}; (v) metallic transition metal dichalcogenide (TMD), such as NbS$_2$ and NbSe$_2$ monolayers with the widely-studied 2H structural phase\cite{TMD}; (vi) $ABA$-stacked few-layer graphene ($ABA$-FLG) where each layer is shifted back and forth by one sublattice; and (vii) $ABC$-stacked few-layer graphene ($ABC$-FLG) where each layer is shifted forward by one sublattice \cite{flg}. Note that the $ABA$ and $ABC$ stacking configurations are chosen here because of their superior stability \cite{neto}.
The analytical RSCs are summarized in Table I.
Remarkably, all RSCs converges to the universal scaling of $\beta = 3/2$ in Eq. (\ref{USR}).
This scaling universality is able to explain three recent experiments of graphene/MoS$_2$ LSH which reported $\beta = 3/2$ scaling in the fitting of the experimental data \cite{ling,guimaraes,behranginia}, rather than the classic diode scaling of $\beta = 2$.

For 2D-material-based VSH, the out-of-plane electron transport can be affected by carrier scattering effects, such as electron-electron interaction and interface inhomogeneities \cite{meshkov, russell, vashaee}. 
In the absence of scattering, the out-of-plane transport conserves the lateral electron momentum, $\mathbf{k}_\parallel$, and depends only on the out-of-plane energy component, $\varepsilon_z$, as customary to the classic Richardson-Dushman model \cite{RD}. 
On the other hand, the presence of scattering effect relaxes $\mathbf{k}_\parallel$-conservation.
Both $\varepsilon_\parallel$ and $\varepsilon_z$ are coupled into the out-of-plane electron transport in this case \cite{vashaee}. 

The RSC across a VSH is generally written as,
\begin{eqnarray}\label{general}
\mathcal{J}_{\star}^{(j)}(k_F, T) &=& \frac{g_{s,v}e}{(2\pi)^2} \int \text{d}^2 \mathbf{k}_\parallel \nonumber \\ 
&\times& \left[ \frac{\lambda j}{L_\perp} \sum_{k_\perp^{(i)}} v_\perp\left[\varepsilon_{\perp}^{(i)}(k_z^{(i)})\right]   f(\mathbf{k}, k_F) \mathcal{T}(\mathbf{k}_\parallel, k_\perp^{(i)}) \right. \nonumber \\
& + & \left. \frac{1}{2\pi} \int \text{d} k_\perp v_\perp\left[\varepsilon_{\perp}(k_\perp)\right] f(\mathbf{k}, k_F) \mathcal{T}(j\mathbf{k}_\parallel, k_\perp) \right], \nonumber \\
\end{eqnarray}
where the subscript `$\star$' emphasizes vertical contact geometry, $\mathbf{k}$ is the total wavevector, $L_\perp$ is the 2D material thickness, $\lambda$ denotes the strength of $\mathbf{k}_\parallel$-non-conserving scattering \cite{meshkov, russell, vashaee}, and the first (second) term denotes the contributions from bound (continuum) electrons with energy below (above) the Schottky barrier.
The $v_\perp[\varepsilon_\perp^{(i)}(k_\perp^{(i)})]$ and $v_\perp[\varepsilon_\perp(k_\perp)]$ are the out-of-plane $z$-directional group velocity for the bound state electron of discrete energy state $\varepsilon_\perp^{(i)}(k_\perp^{(i)})$ and wave vector $k_\perp^{(i)}$, and for the unbound electron of continuous dispersion $\varepsilon_\perp(k_\perp^{(i)})$ and wave vector $k_\perp$, respectively. 
The index $j = 0 (1)$ corresponds to $\mathbf{k}_\parallel$-(non-)conserving model. 
With $j=0$, the first term disappears since the bound state electrons are not energetic enough to overcome $\Phi_{B0}$. 
For $j=1$, the bound state electrons can additionally contribute to the electrical current due to the coupling between $\mathbf{k}_\parallel$ and $k_\perp^{(i)}$. 

Consider the out-of-plane transport from one subband, Eq. (\ref{general}) becomes:
\begin{subequations}\label{j}
	\begin{eqnarray}\label{j0}
	\mathcal{J}_{\star}^{(j=0)} &=& \frac{ e }{ \hbar } k_BT \xi_T  \int_{0}^{\infty} \mathsf{D}(\varepsilon_\parallel) d\varepsilon_\parallel e^{-\frac{\varepsilon_\parallel}{k_BT} },
	\end{eqnarray} 
	\begin{eqnarray}\label{j1}
	\mathcal{J}_{\star}^{(j=1)} &=& \lambda e \left( \frac{\tilde{v}_\perp }{ L_\perp} + \frac{k_BT}{2\pi \hbar}  \right)\int_{\Phi_{B0}}^{\infty} \mathsf{D}(\varepsilon_\parallel) d\varepsilon_\parallel   e^{-\frac{\varepsilon_\parallel - \varepsilon_F}{k_BT}}, \nonumber\\
	\end{eqnarray} 
\end{subequations}
where $\mathsf{D}(\varepsilon_\parallel) $ is the density of state (DOS), $\tilde{v}_\perp \equiv v_\perp\left[\varepsilon_{\perp}^{(1)}(k_z^{(1)})\right]$, $\xi_T \equiv \exp(-\Phi_B/k_BT)$ and the first (second) term in Eq. (\ref{j1}) represents the contribution from bound (continuum) states.

Two important features can be readily seen from Eq. (\ref{j}). 
Firstly, the integral, $\int_{\Phi_{B0}}^\infty d\varepsilon_\parallel (\cdots)$ in Eq. (\ref{j1}) is $\Phi_{B0}$-limited. 
This indicates a strong coupling between the in-plane thermionic process ($\Phi_{B0}$) and the in-plane carrier dynamics ($\varepsilon_\parallel$), thus suggesting the existence of another universal scaling for $\mathcal{J}_{\star}^{(j=1)}$ (see below).
Secondly, Eq. (\ref{j1}) relates the bound and continuum components by $
\mathcal{J}_{\star, \text{continuum}}^{(j=1)} = \eta \mathcal{J}_{\star, \text{bound}}^{(j=1)}$ where $\eta \equiv k_BT  L_\perp/ 2\pi \hbar \tilde{v}_\perp$, $\tilde{v}_\perp = \sqrt{2\varepsilon_\perp^{(1)}/m}$ and $m$ is the free-electron mass.
The upper bound of $\eta$ can be estimated using a finite square well model \cite{vega}.
We first note that since $\eta \propto L_\perp / \sqrt{\varepsilon_\perp^{(1)}} $, $\eta$ can be amplified by a wide and shallow quantum well due to the suppression of electron quantization effect. 
By stretching the 2D material thickness to an exaggerated value of $L_\perp = 5$ nm and considering a weak SBH of $\Phi_{B0} = 0.2$ eV, we obtain $\varepsilon_\perp^{(1)} \approx 4.6$ eV and $\tilde{v}_\perp \approx 1.2 \times 10^6$ m/s. This yields a minimal value of $\eta \approx 0.03$ at room temperature, thus suggesting that the continuum component can be neglected in most cases.
Using graphene with $\Phi_{B0} = 0.5$ eV as an example, we obtain $\varepsilon_\perp^{(1)} \approx 30.5$ eV, and $\tilde{v}_\perp \approx 3.3 \times 10^{6}$ m/s. 
At $T=300$ K and graphene thickness $L_\perp = 0.335$ nm, we obtain $\eta \approx 10^{-3}$, i.e. $\mathcal{J}_{\star, \text{continuum}}^{(j=1)}\ll \mathcal{J}_{\star, \text{bound}}^{(j=1)} $ for graphene.

The $\mathcal{J}_\star^{(j=0,1)}$ is solved for various 2D systems in Table II. 
For graphene (denoted as `Gapless Dirac' in Table II), $\mathcal{J}^{(j=0)}_{\text{Gr}}$ and $\mathcal{J}^{(j=1)}_{\text{Gr}}$ coincides exactly with the Liang-Ang's ($\beta = 3$) \cite{liang} and the Sinha-Lee's \cite{sinha} ($\beta = 1$) VSH model, respectively. 
Here, our generalized model is able to unite the two contsting models under the common physical framework of $\mathbf{k}_\parallel$-conservation: \emph{the Liang-Ang (Sinha-Lee) model belongs to the class of $\mathbf{k}_\parallel$-(non-)conserving continuum (bound) state thermionic transport across a graphene-VSH}.
Importantly, $\mathcal{J}_{\star\text{Gr}}^{(j=1)}$ does not require any arbitrarily-defined parameters, such as the `transit-time' constant, $\tau$, that appears in Sinha-Lee's model with unclear physical meaning \cite{sinha}.

One question then arises: Does VHS respect $\mathbf{k}_\parallel$-conservation in the cross-plane transport?
Here we provide a qualitative estimation using graphene.
The $\mathbf{k}$-non-conserving transport dominates over the $\mathbf{k}_\parallel$-conserving counterpart when the cross-plane transport barrier width is wider than a critical length \cite{russell}, $l_c$, which we estimate as $l_c \approx 1$ nm for $\Phi_{B0} = 0.5$ eV \cite{note}. 
Using typical values of graphene/bulk-semiconductor VSH \cite{tongay}, the width of the nearly triangular-shaped Schottky barrier can be estimated from the depletion width as $W \approx 100$ nm for a reversed bias of 1 V.
For another widely-studied class of Gr/MoS$_2$-thin-film VSH \cite{tomer}, assume that the full thickness of the thin-film (typically $\sim 10$ nm) is depleted, we have $W \approx 10$ nm.
As $W\gg l_c$ in both cases, we thus expect the carrier transport to be predominantly $\mathbf{k}_\parallel$-non-conserving in many VHS samples. 

Under the typical operating condition of $\Phi_{B0} \gg k_BT$, the $\mathcal{J}_\star^{(j=1)}$ in Table II universally converge to $\log\left( \mathcal{J}_{\star}^{(j=1)} / T \right) \propto - 1/T $, i.e. with $\beta = 1$. 
In fact, by expressing the DOS in a general analytic form of $\mathsf{D}(\varepsilon_\parallel) = \sum_\nu a_\nu \varepsilon_\parallel^\nu$, where $\nu \in \mathbb{Z}^{\geq}$ and $a_\nu$ is the expansion coefficient, Eq. (\ref{j1}) yields
\begin{equation}\label{uni2}
\mathcal{J}_{\star}^{(j=1)} \propto \sum_{\nu}T \left[ 1+ \nu \frac{k_BT}{\Phi_{B0}}  \right] \xi_T.
\end{equation} 
This represents another universality of $\beta = 1$, which is supported by recent photo-thermionic experiments in which the measurements are found to be well-reproduced by $|\mathbf{k}_\parallel|$-non-conserving thermionic model \cite{massicotte}.
Using experimental results of graphene-based VSH \cite{tongay, tomer, NR, 2DM}, we found that the $\beta = 1$ scaling provides better fitting with the experimental data as compared to that of the classic $\beta = 2$ scaling.

In summary, we have demonstrated the emergence of a universal scaling exponent $\beta = 3/2$ in lateral Schottky heterostructure, and $\beta = 1$ in vertical Schottky heterostructure with scattering-induced $\mathbf{k}_\parallel$-non-conservation. 
Our findings indicate that the classic diode scaling of $\beta = 2$ for nonrelativistic carrier in bulk material is no longer valid for 2D materials. 
The universal scaling laws developed here shall provide a simple useful tool for the analysis of carrier transport and for the extraction of Schottky barrier height in 2D material Schottky heterostructures, thus paving way towards the design and engineering of novel nanoelectronics, optoelectronics, and spin/valleytronics devices \cite{sv}.
Finally, we remark that graphene-based heterostructures are primarily used to compare with the developed models due to the limited number of reported experimental studies. As the universal scaling laws can be generally applied to broad classes of 2D systems, future experimental verifications of the predicted scaling laws in beyond-graphene systems such as metallic-TMD/semiconducting-TMD heterostructures \cite{LSH} are anticipated.

\begin{acknowledgments}

	This work is supported by A*STAR IME IRG (Grant No. A1783c0011). 

\end{acknowledgments}


\begin{thebibliography}{99}
\bibitem{note} Under thermionic condition ($\varepsilon_\parallel \approx \Phi_{B0}$), $l_c \approx l_0 \log(\eta^{-1/2})$ where $l_0 \equiv \hbar/\sqrt{2m\Phi_{B0}}$ and $\eta \ll 1$ is a dimensionless scattering parameter representing the tunneling rate of $\mathbf{k}$-non-conserving process relative to $\mathbf{k}_\parallel$-conserving process \cite{russell}. We find $l_c \approx 1$ nm for $\Phi_{B0} = 0.5$ eV and a conservative value $\eta = 10^{-3}$.
\end{thebibliography}
\end{document}